# Developing Consistency Among Undergraduate Graders Scoring Open-Ended Statistics Tasks

**Matthew D. Beckman[a]*, Sean Burke[a], Jack Fiochetta[a], Benjamin Fry[a], Susan E. Lloyd[b], Luke Patterson[a], Elle Tang[a]**

a Pennsylvania State University

b Bucknell University

**Authors listed alphabetically*

**ABSTRACT**

Undergraduate graders are frequently important contributors to an instructional team in post-secondary education settings. This study set out to investigate agreement for a team of undergraduate graders as they acquired training and experience for scoring responses to open-ended tasks. Results demonstrate evidence that undergraduate graders can develop the ability to establish and sustain substantial agreement with an instructor, especially when equipped with proper training and a high-quality scoring rubric.

## 1. Introduction and Background

Both the American Statistical Association (ASA) and National Academies of Sciences, Engineering, and Medicine (NASEM) emphasize communication skills as foundational to post-secondary statistics and data science education (ASA Undergraduate Guidelines

Workgroup, 2014; NASEM, 2020). In post-secondary statistics and data science education, such communication skills are expected to be fostered throughout the curriculum (NASEM, 2020), including in introductory courses intended for general education (Gould, 2010). In doing so, research suggests that both statistical literacy and retention of learning improve with frequent writing and communication practice (Graham et al., 2020). In large-enrollment classes, the logistics of implementing and evaluating constructed response tasks jeopardize their use in practice (Garfield & Ben-Zvi, 2008; Woodard & McGowan, 2012). This is not a challenge confined to just a few of the largest research universities, rather, Supiano (2002) reports that a proper majority of undergraduate students enrolled at public institutions in the United States will have at least one large-enrollment Science, Technology, Engineering, or Mathematics (STEM) course.

In post-secondary STEM education and beyond, a popular remedy for economy at scale has long been multiple-choice tasks, calculations, and other easily automated tasks, as a primary vehicle for assessment. Selected-response items (e.g., multiple-choice) elicit an objective form of grading in that there is generally one predetermined correct answer and all graders should assign the same score when labeling student responses with no additional interpretation required on the part of the grader. Open-ended response items (e.g., short-answer response items or constructed response items) elicit a more subjective form of grading. Each student response is worded differently, which can make it more challenging for graders to distinguish correct responses from incorrect responses. The score assigned to a response is based on a myriad of factors including the grader's interpretation of the quality of the response, the grader's interpretation of the content of the response, and can even be influenced by the grader's disposition (Jonsson & Svingby, 2007). As a result, not all graders will necessarily assign the same score when labeling the same short-answer student response.

Rubrics offer a possible remedy to promote greater transparency in the grading process, and making evaluations less subjective by anchoring each score classification to specific response characteristics, exemplars, etc. (Jonsson & Svingby, 2007; Reddy & Andrade, 2009). When developed in this manner, rubrics minimize subjectivity and create structure for the scoring of student performance (Brookhart, 2013). This promotes greater reliability, consistency, and equity in grading across multiple graders, since all graders are using the same rubric to score responses, and all graders are using the same criteria to determine the appropriate score classification (Brookhart, 2013; Jonsson & Svingby, 2007; Ragupathi & Lee, 2020; Reddy & Andrade, 2009; Su, 2020). Even still, the consistency (or agreement) among graders is naturally affected by how many performance categories available and the level of nuance that differentiates those categories when applying the rubric (Jonsson & Svingby, 2007).

Moreover, as the number of graders increases, there is more potential for scoring inconsistencies to emerge (Cathcart & Neale, 2012). In general, training can improve agreement among multiple graders (Jonsson & Svingby, 2007). Rubric training and implementation, in which students actively practice using a rubric, can improve students' awareness and guide students' attention accordingly (Su, 2020). Since undergraduate students often serve as graders on an instructional team, and are involved in the scoring of student performance in post-secondary education settings, it is of interest to formally investigate how rubric training and experience influence the agreement achieved among undergraduate graders scoring responses to open-ended statistics tasks.

## 1.1. ISTUDIO Assessment Tool and Associated Rubrics

As part of an earlier study, an assessment tool called Introductory Statistics Understanding and Discernment Outcomes (ISTUDIO), which consisted entirely of short-answer open-ended tasks, was developed and validated as an instrument for evaluating learning outcomes associated with statistical inference that are typical of a first course in statistics (see Beckman, 2015). As part of this work, detailed rubrics were developed with input from a team of expert statisticians and statistics education researchers, and tested on authentic student responses. The final rubric implemented a three-level scoring approach akin to practices adopted by readers tasked with scoring open-ended tasks for Advanced Placement exams (College Board, 2024). As such, responses to each prompt were evaluated for a holistic score of 0, 1, or 2, such that a score of 0 corresponded to an "incorrect" response, a score of 1 corresponded to a "partially correct" response, and a score of 2 corresponded to an "essentially correct" response. The ISTUDIO rubrics adapted from Beckman (2015), which include a complete statement of each associated task prompt, are provided as supplementary materials.

## 1.2. Previous Scoring Reliability Analysis

As an extension of the 2015 study, a group of researchers performed a scoring reliability analysis on six short-answer response tasks (see Lloyd et al., 2022). These six short-answer response tasks were a subset of the tasks that form the validated ISTUDIO assessment tool developed by Beckman (2015). The Lloyd et al. (2022) study consisted of four human graders, including an experienced statistics instructor and the author of the task prompts and associated rubric (denoted in the study as Rater A), an experienced statistics instructor (denoted in the study as Rater B), a statistics graduate student with experience as an undergraduate mathematics instructor and teaching assistant in statistics (denoted in the study as Rater C), and a statistics graduate student with experience as a teaching assistant in statistics (denoted in the study as Rater D).

In total, the data pool consisted of responses from 1,935 post-secondary students enrolled in introductory statistics courses (see Lloyd et al., 2022 for additional characteristics of the dataset). Initially, the responses from the 1,935 students were assigned to three of the graders (Raters A, B, and C). Each possible combination of the three graders (i.e., the group of all three, and each pair of graders) was randomly assigned a distinct set of 63 students in order to evaluate scoring agreement for that combination. The remaining 1,683 students were split three-ways among the three graders. As a result, the graders were each assigned to review responses from 750 students such that each grader had a unique set of 561 randomly selected students, and each pair of graders shared an intersection of 63 randomly selected students along with a distinct set of 63 randomly selected students shared among all three graders. After the responses had been distributed, but before the scoring took place, the fourth grader (Rater D) joined the study team and was assigned the set of 252 student responses that had been assigned in the intersection of the other three graders, i.e., the total of all 4 sets of 63 responses allocated to each combination of the original three graders.

A brief feasibility study was performed to calibrate the graders' rubric interpretations, discuss ambiguities, and decide how to classify missing and disingenuous responses. For instance, if a student did not attempt a given task, it was labeled as "NA." Each of the four human graders then used the rubric developed in the 2015 study to score the assigned student responses. Agreement among more than two raters was measured using Fleiss' Kappa. Inter-rater agreement between pairs of raters was measured using Quadratic Weighted Kappa (QWK).

QWK values are bound between -1 and 1, such that QWK = 0 represents agreement that would be consistent with random chance. QWK > 0 represents agreement that is better than chance (QWK < 0 is worse than chance), and QWK = 1 is perfect agreement (Fleiss & Cohen, 1973). There are various heuristic interpretations of kappa values to evaluate the strength of agreement (Landis & Koch, 1977). In the Lloyd at al. (2022) study, and in the current study, we adopt Viera and Garrett's (2005) interpretation of the positive kappa values, as follows: 0.01 ≤ kappa ≤ 0.20 is slight agreement; 0.21 ≤ kappa ≤ 0.40 is fair agreement; 0.41 ≤ kappa ≤ 0.60 is moderate agreement; 0.61 ≤ kappa ≤ 0.80 is substantial agreement; 0.81 ≤ kappa ≤ 0.99 is near perfect rater agreement.

The results of the Lloyd et al. (2022) study showed promising evidence of substantial, better-than-chance inter-rater agreement among trained human graders. For instance, when considering a three-way comparison among three of the graders, the Fleiss' Kappa estimate was 0.70; when considering the pairs of graders, the pairwise QWK estimates ranged from 0.79 to 0.83 (Lloyd et al., 2022). For context, one experienced grader in the team evaluated a randomized set of student responses that he had scored seven years prior and achieved an intra-rater agreement of QWK = 0.88 between his implementation of the same rubric to the same responses after approximately 7 years delay, which may inform intuition for a maximum achievable agreement by human graders for the corresponding tasks & rubrics in the study.

## *1.3. Research Questions*

In the Lloyd et al. (2022) study, there were no undergraduate graders on the study team. However, anecdotally, for large classes, or small institutions, undergraduate graders are frequently incorporated into the instructional team. There appears to have been very little research attempting to carefully study the nature of undergraduate graders' development, particularly with respect to consistency (agreement) when scoring open-ended tasks. As such, this research study aims to extend the findings of earlier work related to measuring rater agreement using rubrics, by emulating an instructional team of graders that encompasses a much broader range of teaching experience. In doing so, this research study aims to address the following research question:

*How do agreement with the instructor, self-consistency across time, and group consensus among undergraduate graders change as they receive training and experience scoring open-ended statistics tasks?*

## 2. Methods

### 2.1. Participants

The study team for the current study consisted of an experienced statistics faculty member–Rater A from Lloyd et al. (2022), and hereinafter denoted as the instructor or Rater A–and four undergraduate research assistants (RAs). To provide more context on the study team, all four of the undergraduate RAs (hereinafter denoted collectively as the undergraduate graders, or individually as Raters E, F, G, and H) were students at Penn State University at the time of the study, and had served as undergraduate Learning Assistants for a large introductory statistics course at the University. Eligibility for the Learning Assistants program requires all Learning Assistant candidates to have previously enrolled as a student and succeeded in the same course they intend to support (introductory statistics, in this case). Those accepted to the Learning Assistants program are required to enroll in a pedagogy course, concurrent with their first semester as a Learning Assistant supporting the instructional team. Although all Learning

Assistants were regularly involved in actively supporting students in class and participating in weekly meetings to review class materials and learning objectives with the instructional team, they were *not* involved in grading assignments during their semesters as a Learning Assistant nor had they been formally trained to apply scoring rubrics to open-ended statistics tasks prior to this study.

Among the four undergraduate graders on the study team, two were in their third year and enrolled in either a statistics or data sciences major. Both of these undergraduate graders had served as a Learning Assistant for three semesters prior to the study, though not exclusively for introductory statistics; one of whom had prior research experience in statistics education. The other two undergraduate graders on the study team majored in quantitative fields, but not statistics; both had served as a Learning Assistant for introductory statistics for one semester prior to this study.

### *2.2. Prompts*

The student work to be graded by the study team included student responses to four ISTUDIO prompts and extant data sampled from student responses gathered for a prior study (Beckman, 2015). The four prompts were chosen based on their significance to the broader study. The four prompts had originally been administered to students in two pairs–each pair included a scenario that described a situation that could benefit from data analysis to address a question of interest. The two prompts accompanying each scenario essentially asked a student to explain (a) why statistical inference may or may not be pertinent to the analysis and (b) how the student would actually address the question of interest. In this paper, the prompts are referenced as 2a, 2b, 4a, and 4b to match the original numbering associated with the larger ISTUDIO instrument from which they were adapted.

### *2.3. Progression of Training and Grader Exercises*

The training program and progressive grader exercises for the current study were performed as five sequential grader exercises. The first four grader exercises occurred during the course of one week in May 2024 as the undergraduate graders were participating in an in-person on-boarding workshop at the beginning of a summer research program. The fifth grader exercise occurred approximately 10 weeks later, in August 2024, as the undergraduate graders reached the end of the summer research program. An overview of the training and grader exercise timeline is summarized in Table 1.

| Day 1 | Day 2 | Day 3 | Day 4 | Day 5 | Weeks<br>2 - 10 | Week 10 |
|---|---|---|---|---|---|---|
| Review prompts<br>⬇️<br>Draft solution<br>⬇️<br>Relate solution to rubric<br>⬇️<br>Study team individually grades a set of responses using a minimalist rubric **(grader exercise 1)**<br>⬇️<br>Reflection & discussion | Watch training webinar<br>⬇️<br>In pairs, undergraduate graders make a rubric for a pair of prompts | Divide a new set of responses into 2 equal groups<br>⬇️<br>Undergraduate graders independently grade one group using the pair rubric **(grader exercise 2)**<br>⬇️<br>In pairs, update pair rubric from Day 2<br>⬇️<br>Pairs trade rubrics & use peer rubric to independently grade the second group **(grader exercise 3)***<br><br>*Instructor scores half of each group using each peer rubric | Group discussion about grading from Day 3<br>⬇️<br>Perform grading exercises unrelated to the responses from Days 1 - 3 | Study team individually grades a new set of responses using the expert rubric **(grader exercise 4)** | Summer research program includes grading exercises unrelated to the responses from Days 1-5<br>⬇️<br>Ongoing group discussions about rubric interpretations & grading strategies | Study team individually grades the same responses from Day 5 using the expert rubric **(grader exercise 5)** |

Table 1. Overview of the training and grader exercise timeline in this study.

The data sets associated with each grader exercise included a random sample of responses from 63 students selected from the set of responses provided by the 1,935 students from the 2015 study. A different random sample was drawn for each of the first four grader exercises; the sample associated with grader exercise 4 was reused 10 weeks later for grader exercise 5. Note that the choice of n = 63 was based on a sample size calculation modeled after the Lloyd et al., (2022) study, but the random sampling here was independent of samples selected (or not) for prior studies. The first 4 exercises were completed simultaneously while the group was together for the in-person training workshop, the 5th exercise was completed asynchronously.

A more in depth description of the training follows to expand upon each of the five grader exercises.

### 2.3.1. Day 1 (Grader Exercise 1–Minimalist Rubric)

The study team (consisting of the four undergraduate graders and the instructor) reviewed the prompts they planned to grade and drafted a model solution together. The team discussed each model solution and how it might be interpreted for the purpose of scoring responses on the scale 0, 1, 2, such that a score of 2 represented an "essentially correct" response, a score of 1 represented a "partially correct" response, and a score of 0 represented an "incorrect" response to the prompt. Despite apparent flaws when compared to careful rubric development and implementation, this so-called "minimalist rubric" was implemented as a sort of baseline that was intended to resemble a common strategy encountered by student graders (undergraduate or otherwise) wherein the instructor provides a model solution, discusses general grading priorities with the graders, and then the graders apply this strategy to a set of student work as best they can.

All members of the study team individually graded the first set of student responses using the minimalist rubric (the model solution and verbal instruction). The study team followed the grader exercise with a brief discussion reflecting on the task.

### 2.3.2. Day 2 (Training Webinar)

The study team watched a training webinar together to discuss the use, development, and implementation of rubrics for scoring short-answer tasks (Ziegler, 2024). This workshop was specifically commissioned for this project and was tailored to the audience's needs and objectives, but a recording is publicly available at https://www.causeweb.org/cause/webinar/classifies-webinar.

Following the webinar, the undergraduate graders were split into two pairs. Within each pair, the undergraduate graders co-created a rubric associated with one pair of prompts, e.g., one team of students developed a rubric for ISTUDIO prompts 2a & 2b, while the other pair developed a rubric for prompts 4a & 4b. As a result, two rubrics were created (one rubric created by each pair of undergraduate graders). We hereafter refer to these rubrics as the "pair rubrics" since these rubrics were co-created by pairs of the undergraduate graders.

### 2.3.3. Day 3 (Grader Exercise 2–Pair Rubric and Grader Exercise 3–Peer Rubric)

A new set of student responses was provided to each pair of undergraduate graders and split into two halves. Within each pair, each undergraduate grader independently graded half of the assigned student responses using the "pair rubric" they helped develop. Note that students graded different subsets of this shared data set. The undergraduate graders were encouraged to discuss any student responses that were challenging to grade and to refine their pair rubrics accordingly. After one half of the assigned student responses had been graded, the undergraduate graders worked within their respective pairs to refine any necessary modifications and finalize their pair rubrics.

Since the instructor did not participate in the creation of the pair rubrics, the instructor does not have any data for grader exercise 2. For this reason, along with the fact that the undergraduate graders were encouraged to discuss their scoring within their pairs, further analysis of grader exercise 2 was deemed out of scope for the present study.

Shortly after completing grader exercise 2, the two pairs of undergraduate graders were asked to trade their refined rubrics to simulate implementing a rubric created by someone else. Each undergraduate grader independently graded the third set of responses (i.e., the other half of the responses from grader exercise 2) with the rubric created by the other pair of undergraduate graders. We hereafter refer to these rubrics as the "peer rubrics" since the rubrics being implemented were created by the other pair of undergraduate graders.

During the live training workshop, the instructor scored a subset of each pair's third set of responses with each of the rubrics developed by the pairs of undergraduate graders (in particular, the instructor scored approximately 30 responses using the each of the rubrics developed by the two pairs of undergraduate graders. This compromise was decided in the moment since the entire group was working simultaneously in person, and it would have delayed subsequent activities planned for the training program if the instructor were to grade both data sets before the group could continue. Thus, when performing the rater agreement analyses, the instructor has two data points for grader exercise 3, one for each sample scored using each of the peer rubrics. In this way, the data scored by the instructor is substantively different from the data scored by the undergraduate graders, which will be considered when interpreting the rater agreement analyses for this grader exercise.

#### 2.3.4. Day 4 (Training Continued)

The study team had an informal discussion of grader exercises 2 and 3 from the previous day, and then spent the day working with expert rubrics and tasks associated with the larger research objectives of the summer research program (i.e., not associated with ISTUDIO).

#### 2.3.5. Day 5 (Grader Exercise 4–Expert Rubric, Part 1)

The instructor provided the undergraduate graders with the rubric that had been rigorously developed in the 2015 study to accompany the ISTUDIO tasks for research purposes. This rubric, denoted as the “expert rubric”, is provided as a supplementary material. The expert rubric was more detailed than the rubrics discussed or implemented previously by the study team. Using this rubric, all members of the study team individually graded the fourth set of responses.

#### 2.3.6. Weeks 2 through 10

During weeks 2 through 10, the study team participated in a summer research program together. The work involved a considerable amount of time using rubrics to score student responses to introductory statistics prompts. The study team frequently discussed rubric interpretations and strategies to drive consistency. The relevant tasks and rubrics were pertinent to a larger, related, research project, but the prompts and rubrics were not associated with ISTUDIO during this period. The multi-week summer research program took place in between grader exercises 4 and 5 to provide the opportunity to evaluate whether undergraduate graders were able to sustain consistency in assessing open-ended responses using the expert rubric.

#### 2.3.7. Week 10 (Grader Exercise 5–Expert Rubric, Part 2)

After a 10-week elapsed period of grading a variety of different tasks and implementing associated rubrics (generally quite distinct from those encountered during the training week), each grader was given the fifth set of responses, which was identical to the set of responses evaluated in grader exercise 4. As in grader exercise 4, all members of the study team again used the same expert rubric to grade the fifth set of responses, but now after many weeks of experience participating in a statistics education research program.

### *2.4. Rater Agreement Analysis*

For pairwise (inter-rater agreement), we estimate the scoring reliability using QWK. We estimate the inter-rater agreement between each undergraduate grader and the instructor (i.e., Rater A) for grader exercises 1, 3, 4, and 5. Since grader exercise 4 and grader exercise 5 utilize the same set of student responses using the same rubric, we also use QWK to estimate the self-consistency (intra-rater agreement) between the scores provided by each undergraduate grader in grader exercise 4 and grader exercise 5.

Additionally, for the analysis of agreement among the entire group of graders, we use Gwet's $AC_2$, which is appropriate for use when multiple graders are all scoring an identical set. Gwet's $AC_2$ is selected as the measure of agreement since it allows for ordinal weights and measures agreement in a manner that allows for some continuity in interpretations for the previous and present studies (Gwet, 2014). In particular, it uses the same percent agreement as Fleiss' Kappa (Gwet, 2014), as used in the 2022 study discussed previously. There is some concern that the absolute measure of $AC_2$ can become inflated for unbalanced data sets (Tran et al., 2018), but since the marginal balance is unchanged across the various grader exercises being investigated, the relative comparisons are not jeopardized. Analysis of group agreement will be reported only for grader exercises 1, 4, and 5, since all graders had scored an identical set of student responses for each of those grader exercises. By contrast, graders had scored different subsets of a shared data set for grader exercises 2 and 3, which precludes analysis of group agreement.

## 3. Results

### *3.1. Preliminary Scoring Analysis*

Figure 1 illustrates, for each prompt across each of the five grader exercises, the mean score given by each of the five graders on the study team. There was no missingness or aberrations from the data collection plan associated with Exercises 1, 4, or 5. Note that the instructor (Rater A) is absent from grader exercise 2, for reasons previously described. Also, note that the instructor (Rater A) has two data points for grader exercise 3 since the instructor had used the rubric created by each of the two pairs of undergraduate graders to score a subset of each sample. Also note that Rater G is absent from grader exercise 3, because Rater G had inadvertently scored a different data set. Recall that the rubrics in grader exercise 2 were still being refined by the undergraduate graders who were scoring in tandem. Also, each grader was instructed to be as faithful as possible to the provided rubric for grader exercise 3–even if the grader might not agree with a score prescribed by the rubric. Therefore, comparisons of the variability of mean scores among the graders are perhaps most usefully interpreted for grader exercises 1, 4, and 5, although all grader exercise results are shown for completeness.

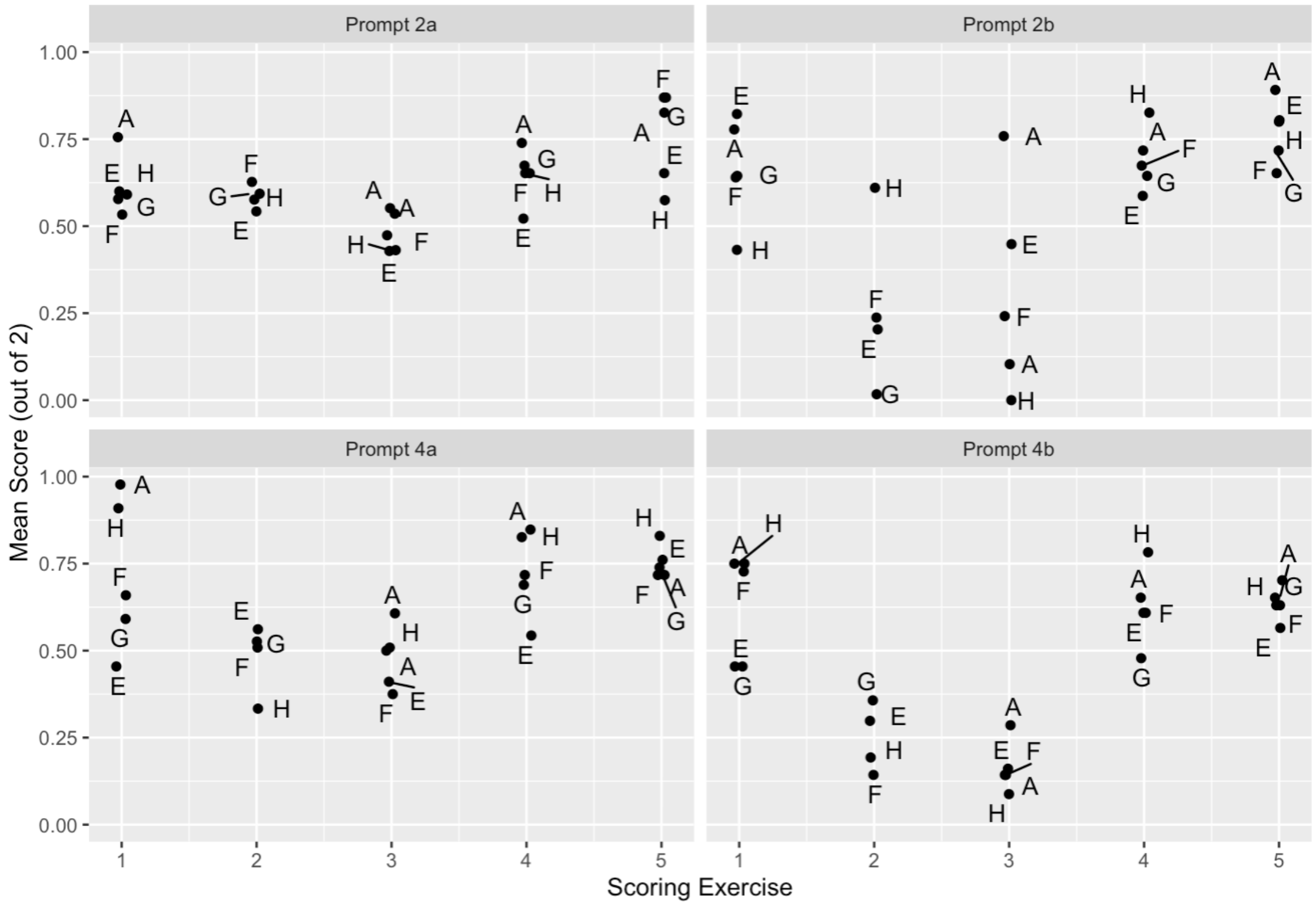

Prompt 2a
Prompt 2b
Prompt 4a
Prompt 4b
Mean Score (out of 2)
Scoring Exercise

*Figure 1.* Mean scores calculated for each of 5 study team members evaluating samples of student responses to ISTUDIO prompts (i.e., 2a, 2b, 4a, 4b) with 4 different rubric conditions (i.e., grader exercises 1, 2, 3, 4). Grader exercise 5 replicates the conditions of grader exercise 4 approximately 10 weeks later.

By inspection, when comparing solely grader exercises 1, 4, and 5, it appears that the mean scores associated with Prompt 2a change very minimally across the grader exercises. By comparison, again only considering grader exercises 1, 4, and 5, the mean scores associated with Prompt 2b, Prompt 4a, and Prompt 4b appear somewhat less variable from grader to grader with each successive grader exercise.

The mean scores associated with each grader appear to fluctuate depending on the grader exercise. For example, consider Prompt 2b. Rater H initially had the lowest mean score (grader exercise 1), but after a week of rubric training, Rater H had the highest mean score (grader exercise 4); whereas, Rater E initially had the highest mean score (grader exercise 1), but after a week of rubric training, Rater E had the lowest mean score (grader exercise 4). Similarly, when considering Prompt 2a, Raters F and G initially had some of the lowest mean scores (grader exercise 1), but after extensive rubric training, these same raters had the highest mean scores (grader exercise 5). Across all prompts, the instructor (Rater A) tends to have a mean score that is slightly above average, relative to the undergraduate graders.

## *3.2. Participant Reflections about Progression of Training and Grading*

Accompanying the first four grader exercises during the training week, the study team had informal discussions about each grader exercise. Grader exercise 1 (using a minimalist rubric), took approximately one hour for all graders to complete. Graders described strategies such as creating a “mental rubric” to try and remember guidelines they had intended to implement, while other graders discussed frequently referring back to responses they had previously scored and then using comparative judgments (i.e., is a given response higher, lower, or similar quality to a previously scored one).

Grader exercises 2 (Pair Rubric) and 3 (Peer Rubric) were administered in close succession on the same day. Overall, these grader exercises took a similar amount of time as the first grader exercise to complete. The graders reported that their grading felt more “strict” when using a rubric, as there was less room to give students the “benefit of the doubt.” For Prompts 2b and 4b, the mean scores for grader exercises 2 and 3 do generally appear quite a bit lower by inspection of the comparison with grader exercise 1, according to Figure 1. However, the graders also reported that it was easier to make decisions on scores with a rubric and they did not rely as heavily on comparative judgments with other responses.

Grader exercise 4 (Expert Rubric, Part 1) took approximately 60-75 minutes for the undergraduate graders to complete, but it took the instructor noticeably less time to complete, which is understandable as the instructor had been so instrumental in creating the rubric. Grader exercise 5 (Expert Rubric, Part 2) was completed asynchronously and was not timed. Also, there was not a dedicated group discussion to reflect on the fifth grader exercise.

### *3.3. Pairwise Agreement for Each Undergraduate Grader with the Instructor*

The pairwise agreement for each undergraduate grader with the instructor (Rater A) is reported in Table 2. As noted in the preliminary scoring analysis, Rater G had inadvertently scored a different data set than the other graders for grader exercise 3, so there was not an intersecting set of responses available to evaluate inter-rater agreement in that case.

| Raters | Exercise 1 | Exercise 3 | Exercise 4 | Exercise 5 |
|---|---|---|---|---|
| A & E | 0.46 (0.35, 0.58) | 0.54 (0.42, 0.67) | 0.57 (0.47, 0.67) | 0.58 (0.49, 0.67) |
| A & F | 0.61 (0.50, 0.71) | 0.63 (0.51, 0.74) | 0.72 (0.64, 0.79) | 0.78 (0.71, 0.85) |
| A & G | 0.63 (0.55, 0.72) | NA | 0.73 (0.66, 0.80) | 0.73 (0.66, 0.81) |
| A & H | 0.72 (0.65, 0.80) | 0.60 (0.48, 0.72) | 0.71 (0.63, 0.78) | 0.68 (0.59, 0.78) |

Table 2. Pairwise agreement for each undergraduate grader with the instructor (Rater A) as measured with Quadratic Weighted Kappa; 95% confidence intervals accompany each estimate.

For Raters E, F, and G, the results suggest evidence of an improvement in pairwise agreement between undergraduate graders and the instructor (Rater A) from grader exercise 1 on Day 1 to grader exercise 4 on Day 5. However, the agreement between one of the undergraduate graders (Rater H) and the instructor (Rater A) was largely unchanged over this same time period. For many of the rater pairs, there was negligible evidence of improved agreement between each undergraduate grader and the instructor in Week 10 (grader exercise 5), following many weeks of additional experience refining and applying rubrics for the purpose of scoring other statistics tasks. Only Rater F showed evidence of improved agreement with the instructor after many weeks of additional experience applying rubrics. However, the improved agreement between Rater F and the instructor from grader exercise 4 to grader exercise 5 was not as noticeable as their improved agreement from grader exercise 1 to grader exercise 4 attributed to the training week.

Considering the QWK point estimates, if we adopt Viera and Garrett's (2005) interpretation of the positive kappa values, "moderate agreement" is observed between Rater A and Rater E, and "substantial agreement" is observed between Rater A and Raters F, G, and H throughout the study (with the exception of the agreement between Raters A and H for grader exercise 3 which falls slightly below Viera and Garrett's (2005) threshold for "substantial agreement").

Considering the 95% confidence intervals for the QWK estimates, "fair" to "moderate agreement" is observed between Rater A and Rater E on Day 1, but this agreement improves to a value on Day 3 that could be described as "moderate" to "substantial agreement," which is sustained throughout the study. Similarly, "moderate" to "substantial agreement" is observed between Raters A & F and Raters A & G on Day 1; this agreement improves to a value on Day 5 that would be described as "substantial," which was then sustained until Week 10.

Rater A and Rater H achieve "substantial" agreement from Day 1, and then the agreement does not appreciably change during the course of the 10 weeks; the slight decline of the QWK estimate by 0.04 over the course of the study does not represent compelling evidence of a statistically–or practically–meaningful effect.

### *3.4. Self-Consistency for Each Grader on the Study Team*

Since grader exercise 4 and grader exercise 5 utilize the same data set scored approximately 10 weeks apart, we can analyze the self-consistency of each grader (i.e., intra-rater agreement) as shown in Table 3.

| Raters | QWK | 95% CI |
|---|---|---|
| A | 0.82 | (0.76, 0.88) |
| E | 0.57 | (0.46, 0.68) |
| F | 0.74 | (0.67, 0.82) |
| G | 0.66 | (0.56, 0.76) |
| H | 0.74 | (0.67, 0.81) |

Table 3. Intra-rater agreement (self-consistency) for each grader on the study team as measured with QWK while scoring the same set of student responses on two separate occasions approximately 10 weeks apart.

The 95% confidence interval for the self-consistency of the instructor (Rater A) is (0.76, 0.88). According to Viera and Garrett (2005), this represents "substantial" to "near perfect" agreement.

Raters F and H have the highest self-consistency among the undergraduate graders with 95% confidence intervals of (0.67, 0.82) and (0.67, 0.81), respectively. These intervals suggest that Raters F and H also have "substantial" to "near perfect" agreement with themselves, even after almost 10 weeks have elapsed; however, their intra-rater agreement is not as strong as that of the instructor. The 95% confidence interval for the self-consistency of Rater G is (0.56, 0.76), while the 95% confidence interval for Rater E is (0.46, 0.68). These intervals suggest that Raters G and E have weaker ("moderate" to "substantial") intra-rater agreement with themselves than do the instructor and the other undergraduate graders.

### *3.5. Agreement Among Entire Study Team*

Gwet's $AC_2$ was computed as a measure of consensus among the entire study team for grader exercises 1, 4, and 5 (see Table 4). For grader exercise 1 (on Day 1 using a minimalist rubric), Gwet's $AC_2$ (0.688) is comparable to the Fleiss' Kappa among three raters (0.70) observed in the Lloyd et al. (2022) study. This suggests that, when using a minimalist rubric, an instructional team consisting of multiple undergraduate graders and an instructor can expect to achieve similar agreement as that of a team of more experienced human raters, even before undergoing extensive training.

| Date (Exercise) | Rubric Description | $AC_2$ | 95% CI |
|---|---|---|---|
| Day 1 (Ex 1) | Solution with Verbal Instructions | 0.688 | (0.63, 0.74) |
| Day 5 (Ex 4) | Expert Rubric, Part 1 | 0.784 | (0.75, 0.82) |
| Week 10 (Ex 5) | Expert Rubric, Part 2 | 0.778 | (0.74, 0.81) |

Table 4. Group agreement among four undergraduate graders and one instructor, as measured with Gwet's $AC_2$; 95% confidence intervals accompany each estimate.

By inspection of the 95% confidence intervals associated with each grader exercise, there is evidence of stronger agreement among the entire study team for grader exercise 4 (on Day 5 with the Expert Rubric) when compared to grader exercise 1 (on Day 1 with a minimalist rubric). This increase appears to be sustained as much as 10 weeks later, as there is also evidence for stronger agreement among the entire study team for grader exercise 5 (in Week 10 with the Expert Rubric) as compared to grader exercise 1 (on Day 1 with a minimalist rubric), and this evidence is comparable to that from grader exercise 4 (on Day 5 with the Expert Rubric). This provides preliminary evidence that, with the proper training, an instructional team of primarily undergraduate graders can expect to achieve stronger agreement than that which is achieved among a team of experienced human raters without this level of rubric training. Also, the training required to achieve this improved agreement can be implemented in a relatively short period of time (one week) in the event that an entire 10 week program is not feasible to replicate.

## 4. Discussion

*Recall the research question: How do agreement with the instructor, self-consistency across time, and group consensus among undergraduate graders change as they receive training and experience scoring open-ended statistics tasks?*

This study demonstrates evidence that undergraduate graders can develop the ability to establish and sustain substantial agreement with an instructor when scoring open-ended statistics tasks. Several undergraduate graders made noticeable gains in their agreement with the instructor (Rater A) when using a quality rubric at the end of their first week, which appears to persist for the duration of the 10 week program. While the use of a minimalist rubric, consisting only of a model solution and verbal instruction, more often resulted in weaker agreement with the instructor, several undergraduate graders improved their agreement with the instructor after a few days of training and sustained those gains for many weeks thereafter.

When compared to a similar study evaluating agreement among more highly trained study team members, Lloyd et al. (2022) observed inter-rater agreements that were only slightly higher than those observed for many of the trained undergraduate graders using an expert rubric. Group consensus among all five graders (four undergraduate graders and one instructor) follows a similar trend, including a noticeable improvement when using a quality rubric at the end of the first week.

It is worth noting that Rater H, who was the only undergraduate grader that did not have improved agreement with the instructor, was the most experienced of the undergraduate graders. Rater H had completed more advanced statistics coursework than the other undergraduate graders, and had also been part of a different statistics education study team for a full year prior to joining this study team and participating in the study. There is precedent in the literature suggesting that evaluative judgments are correlated with prior expertise with the subject matter– both statistics and statistics education research in this case. For example, Boud, Lawson, and Thompson (2013) found a similar effect in the context of self-evaluation of student work and agreement with an expert/tutor's evaluation of the same student work. Similarly, there is precedent in the literature about expert schemas, suggesting that students' cognitive structures become more aligned with that of an expert as students develop deeper, more connected understanding (Schau & Mattern, 1997). Since Rater H had the most experience with statistics coursework and research, Rater H could have entered the study with an expert schema that was more aligned with that of the instructor, which would be consistent with results observed in this study.

From our analysis of intra-rater agreement, we see that the instructor (Rater A) has the highest self-consistency, and the self-consistency of each undergraduate grader tends to be similar to the inter-rater agreement between that undergraduate grader and the instructor after one week of training. Rater G was an exception, since the intra-rater agreement for Rater G was estimated to be QWK = 0.66 with a 95% confidence interval of (0.56, 0.76), and yet the inter-rater agreement between Rater G and Rater A was estimated to be QWK = 0.73 for both grader exercises 4 and 5. This was an unexpected result, as it suggests that Rater G has stronger agreement with Rater A than with themself. However, since the inter-rater agreement between Raters A and G is within the 95% confidence interval for Rater G's intra-rater agreement, this could simply be an artifact of randomness.

Analysis of group agreement suggests that there is evidence of improved agreement among the five graders on the study team following a training webinar and a series of five grader exercises. The improvement appears to have been sustained as much as 10 weeks later.

### *4.1. Limitations*

Analysis of quantitative results associated with grader exercises 2 and 3 are limited because of the manner in which the student responses for these two grader exercises were scored. The available data for these two grader exercises is not commensurate with the type of data used for the other comparisons reported in this paper. For example, for grader exercise 2, the instructor (Rater A) did not create a rubric; for grader exercise 3, the instructor only scored partial data sets using each of the two rubrics generated by the pairs of undergraduate graders in order to sample each and facilitate discussion in the training on Day 4.

Moreover, on Day 3, the undergraduate graders were invited to split the data set in half so that they could evaluate one half of the student responses with the rubric that they had helped create (ie., grader exercise 2; pair rubric) and the other half of the student responses with the rubric that others had created (ie., grader exercise 3; peer rubric). Due to the manner in which the student responses for grader exercises 2 and 3 were scored, this limits the analyses that can be performed on these two grading exercises. However, it should be noted that participating in these two grader exercises may still have influenced the observed changes in agreement between grader exercise 1 on Day 1 and grader exercise 4 on Day 5. Alternatively, it is possible that the observed changes may be attributed to the quality of the rubric used in grader exercise 4 (the Expert Rubric).

Due to the circumstances of the larger project within which this study was conducted, several interventions are confounded here, which may warrant further investigation. For example, no attempt was made to investigate the minimum training intervention necessary to achieve sufficient grading performance, such as isolating the effect associated with the training webinar or investigating scoring performance with a minimalist rubric (i.e., model solution and verbal instruction) after graders had already demonstrated competency with the use of an expert rubric.

Furthermore, the observed changes in agreement could be due to the nature of an immersive summer research program experience. The undergraduate graders were working with the tasks and rubrics extensively over the course of these 10 weeks, which naturally could have led to a deeper understanding of the statistical concepts being assessed and a greater confidence in applying the rubric to student responses. Alternatively, the undergraduate graders could have become more likeminded in their approach to grading as a result of spending these 10 weeks together in an immersive research experience.

Nevertheless, the findings of this study illustrate that the undergraduate graders were able to achieve, and maintain, improved agreement with the instructor (and as a group) as a result of using a quality scoring rubric and participating in a relatively short summer research program.

### *4.2. Implications for Teaching*

The results of this study demonstrate evidence that undergraduate graders can develop the ability to establish and sustain substantial agreement with an instructor, especially when equipped with proper training and a high-quality scoring rubric. Of course, it is important to note that the undergraduate graders already had experience with the relevant course material and had themselves succeeded as students previously in the relevant course. Although all four of the undergraduate graders on this study team had chosen quantitative subjects as an academic major, they had not all chosen to major in statistics or data science, which matches intuition that a broader pool of quantitatively oriented undergraduates can perform well as graders for an introductory statistics course when provided with the proper training and tools.

Although it may not be feasible in all circumstances for an instructional team to train undergraduate graders according to the 10 week program discussed in this study, unless this training were to be integrated throughout a course or degree program, the noticeable gains in agreement were primarily observed from grader exercise 1 (on Day 1) to grader exercise 4 (on Day 5). This agreement was then sustained throughout the summer research program until Week 10. Thus, these results suggest that a one week training should be a sufficient amount of time to improve undergraduate graders' agreement to a level that resembles that of more expert graders, and this agreement can be expected to be sustained throughout the duration of the semester.

## Supplementary Materials

The supplementary materials include (1) the ISTUDIO rubrics with verbatim task prompts, (2) de-identified ISTUDIO data with scoring labels, (3) R code to reproduce the results presented in this manuscript.

## Data Availability Statement

The data that support the findings of this study are openly available in Figshare at http://doi.org/10.6084/m9.figshare.27327774

## Acknowledgements

We're grateful to the US National Science Foundation for funding this research (NSF DUE-2236150: Project CLASSIFIES), as well as input from Laura Ziegler, Dennis Pearl, and Rebecca Passonneau.

## Ethics and Disclosure Statement

No potential conflict of interest was reported by the author(s). Study design and data collection were reviewed and deemed IRB exempt by the Penn State University Office for Research Protections (Study ID # 21834).